\documentclass[twocolumn,showpacs,preprintnumbers,amsmath]{revtex4}
\usepackage{amssymb}

\usepackage{graphicx}
\usepackage{dcolumn}
\usepackage{bm}

\begin{document}

\title{Theory of spectrum in qubit-Oscillator systems in the ultrastrong coupling
regime}
\author{Qing-Hu Chen$^{1,2}$, Lei Li$^{3}$, Tao Liu$^{3}$, and Ke-Lin Wang$^{4}$}

\address{
$^{1}$ Center for Statistical and Theoretical Condensed Matter
Physics, Zhejiang Normal University, Jinhua 321004, P. R. China  \\
$^{2}$ Department of Physics, Zhejiang University, Hangzhou 310027,
P. R. China \\
$^{3}$Department of Physics, Southwest University of  Science and Technology, Mianyang 621010, P.  R.  China\\
$^{4}$Department of Modern Physics, University of  Science and
Technology of China,  Hefei 230026, P.  R.  China
 }
\date{\today}

\begin{abstract}
Recent measurement on an LC resonator magnetically coupled to a
superconducting qubit[arXiv:1005.1559] shows that the system
operates in the ultra-strong coupling regime and crosses the limit
of validity for the rotating-wave approximation of the
Jaynes-Cummings model. By using extended bosonic coherent states, we
solve the Jaynes-Cummings model exactly without the rotating-wave
approximation. Our numerically exact results for the spectrum of the
flux qubit coupled to the LC resonator are fully consistent with the
experimental observations. The smallest Bloch-Siegert shift obtained
is consistent with that observed in this experiment. In addition,
the Bloch-Siegert shifts in arbitrary level transitions and for
arbitrary coupling constants are predicted.
\end{abstract}
\pacs{42.50.Pq, 03.65.Ge, 85.25.Cp, 03.67.Lx}
 \maketitle

\section{Introduction}
The Jaynes-Cummings (JC) model\cite{JC} describes the interaction of
a two-level atom with a single bosonic mode, which is fundamental
model in quantum optics. Recently, the JC model is also closely
related to condensed matter physics. It can be realized in some
solid-state systems, such as one Josephson charge qubit coupling to
an electromagnetic resonator \cite {Wallraff}, the superconducting
quantum interference device coupled with a nanomechanical
resonator\cite{squid}, and the  LC resonator magnetically coupled to
a superconducting qubit\cite{exp}. In conventional quantum optics,
the coupling between the "natural" two-level atom and the single
bosonic mode is quite weak, the rotating-wave approximation (RWA)
has been usually employed. With the advent of circuit quantum
electrodynamics (QED), on-chip superconducting qubits (the
"artificial " two-level atoms) could be engineered to interact very
strongly with   oscillators (cavities)\cite
{Wallraff,squid,Chiorescu,exp,Schuster,Deppe,Fink,Hofheinz}, RWA can
not describe well the strong coupling regime\cite{liu}, so the
studies to the JC model without RWA is highly called for.

However, it is more difficult to solve the JC model without RWA than
with RWA. In the absence of RWA, due to the presence of the
counter-rotating terms, the photonic number is not conserved, so the
photonic Fock space has infinite dimensions. The standard
diagonalization procedure (see, for example, Ref. \cite{exp}) is the
first candidate, which is to apply a truncation procedure
considering only a truncated number of photons. Typically, the
convergence is assumed to be achieved if the numerical results are
determined within very small relative errors. Within this method,
one has to diagonalize very large, sparse Hamiltonian in strong
coupling regime. Furthermore, the calculation might become
prohibitive for higher excited states where more photons should be
involved.

Fortunately, several non-RWA approaches\cite
{chenqh,zheng,liu,Liutao,Amico,Yuyu} has been recently proposed in a few
contexts. Especially, by using extended bosonic coherent states, three of
the present authors and a collaborator have solved the Dicke model without
RWA exactly in the numerical sense\cite{chenqh}. The JC model is just
special Dicke model with only one two-level atom.

Recently,  the spectrum for an LC resonator magnetically coupled to
a superconducting qubit was measured experimentally.  A 50 MHz
Bloch-Siegert shift when the qubit is in its symmetry point was
observed, which clearly shows that the system enter the ultra-strong
coupling regime. Therefore   JC model with RWA is invalid to
describe this strong coupling system. In this paper, we numerically
solve JC model without RWA exactly. Based on the some key data drew
from the spectrum\cite{exp}, we obtain a fit of the experimental
parameters. All spectrum line can then be calculated. The
Bloch-Siegert shifts in arbitrary level transitions and in a wide
range of the coupling parameters can also be estimated.

The paper is organized as follows. In Sec.II, the numerically exact
solution to the JC model is proposed in detail. The numerical
results and discussions are given in Sec.III. The brief summary is
presented finally in the last section.

\section{Model}

The interaction between the flux qubit and the LC resonator in the
experiment \cite{exp} is described by
\begin{equation}
H_{int}=\hbar g(a^{\dagger }+a)\sigma _z
\end{equation}
where $a^{\dagger }$, $a$ are the photon creation and annihilation
operators in the basis of Fock states of the LC resonator, $g$ is
the   flux qubit-cavity coupling constant. The RWA has not been
employed here. The effective Hamiltonian for the flux qubit can be
written as the standard one for a two-level system
\begin{equation}
H=-\left( \epsilon \sigma _z+\Delta \sigma _x\right) /2
\end{equation}
where $\Delta $ and $\epsilon $ are is the tunneling coupling
between the two persistent current states and the transition
frequency of the flux qubit.  $\epsilon=I_p (\Phi-\Phi_0/2) $ with
$I_p$ the persistent current in the qubit loop, $\Phi$ an externally
applied magnetic flux, and  $\Phi_0$ the flux quantum. In the above
two equations, the Pauli matrix notations $\sigma _k(k=x,y,z)$ $\ $
are used in the basis of the two persistent current states. Then the
Hamiltonian for the whole system  reads
\begin{eqnarray}
H &=&-\left( \epsilon \sigma _z+\Delta \sigma _x\right) /2+\hbar
\omega
_r\left( a^{\dagger }a+\frac 12\right)  \nonumber \\
&&+\hbar g(a^{\dagger }+a)\sigma _z
\end{eqnarray}
where $\omega _r$ is the cavity frequency. For convenience, we
denote
\[
\hbar \omega _q=\sqrt{\epsilon ^2+\Delta ^2},\tan \theta =\Delta /\epsilon
\]
Then the final Hamiltonian is ( $\hbar \;$is set to unity)
\begin{eqnarray}
H &=&-\frac{\omega _q}2\left[ \cos (\theta )\sigma _z+\sin (\theta
)\sigma
_x\right] +\omega _r\left( a^{\dagger }a+\frac 12\right)  \nonumber \\
&&+g\left( a^{\dagger }+a\right) \sigma _z
\end{eqnarray}

By introducing the new operators
\begin{equation}
A=a+\alpha ,B=\alpha -a,\alpha =g/\omega _r
\end{equation}
we have
\begin{equation}
H=\left(
\begin{array}{cc}
\omega _r\left( A^{\dagger }A-\alpha ^2\right) +\epsilon _{-} &
-\omega
_q\sin (\theta )/2 \\
-\omega _q\sin (\theta )/2 & \omega _r\left( B^{\dagger }B-\alpha
^2\right) +\epsilon _{+}
\end{array}
\right)
\end{equation}
where $\epsilon _{\pm }=\left( \omega _r\pm \omega _q\cos \theta
\right) /2$.  Note that the linear term for the original  bosonic
operator $a^{\dagger }(a)$ is removed, and only the  number
operators $A^{+}A$ and $B^{+}B$ are left. Therefore the wavefunction
can be expanded in terms of these new operators as
\begin{equation}
\left| {}\right\rangle =\left(
\begin{array}{l}
\left| \varphi _1\right\rangle  \\
\left| \varphi _2\right\rangle
\end{array}
\right) =\left(
\begin{array}{l}
\sum_{n=0}^{N_{tr}}c_n\left| n\right\rangle _A \\
\sum_{n=0}^{N_{tr}}d_n\left| n\right\rangle _B
\end{array}
\right)
\end{equation}
For $A$ operator, we have
\begin{eqnarray}
\left| n\right\rangle _A &=&\frac{A^n}{\sqrt{n!}}\left| 0\right\rangle _A=%
\frac{\left( a+\alpha \right) ^n}{\sqrt{n!}}\left| 0\right\rangle _A \\
\left| 0\right\rangle _A &=&e^{-\frac 12\alpha ^2-\alpha
a^{+})}\left| 0\right\rangle _a.
\end{eqnarray}
 $B$ operator has the same properties. Inserting Eqs. (6) and (7) into the Schr$\stackrel{..}{o}$ dinger equation, we have
\begin{eqnarray}
\left[ \epsilon _{-}+\omega _r\left( m-\alpha ^2\right) \right] c_m
&&
\nonumber \\
-\frac{\omega _q\sin (\theta )}2\sum_nD_{mn}d_n &=&Ec_m \\
\left[ \epsilon _{+}+\omega _r\left( m-\alpha ^2\right) \right] d_m
&&
\nonumber \\
-\frac{\omega _q\sin (\theta )}2\sum_nD_{mn}c_n &=&Ed_m
\end{eqnarray}
where
\[
D_{mn}=\exp (-2\alpha ^2)\sum_{k=0}^{\min
[m,n]}(-1)^{-k}\frac{\sqrt{m!n!} (2\alpha
)^{m+n-2k}}{(m-k)!(n-k)!k!}
\]
In principle, all eigenvalues and eigenfunctions can be obtained in
Eqs. (10) and (11). As before, to obtain the true exact results, the
truncated number $N_{tr}$ should be taken to infinity. Fortunately,
it is not necessary. It is found that finite terms in state (7) are
sufficient to give very accurate results with a relative errors less
than $10^{-5}$ in the whole parameter space. We believe that we have
exactly solved the JC model numerically. The numerical results are
given in the next section.

\textsl{Solutions in the  symmetry point}.-- The spectrum in the
symmetry point ($\epsilon=0$ ) is particularly interesting in
experiments. In this case, Eq. (3) becomes
\begin{equation}
H_0=-\frac \Delta 2\sigma _x+\hbar \omega _r\left( a^{\dagger
}a+\frac 12\right) +g\left( a^{\dagger }+a\right) \sigma _z
\end{equation}
Associated with this  Hamiltonian is a conserved parity $\Pi$, such
that $\left[ H_0,\Pi \right] =0$, which is given by
\begin{equation}
\Pi =e^{i\pi \sigma _y/4}e^{i\pi \widehat{N}}e^{-i\pi \sigma _y/4},
\widehat{N}=a^{\dagger }a+\sigma _z/2+1/2,
\end{equation}
where $\widehat{N}$ is the excitation number operator. $\Pi$ has two
eigenvalues $\pm 1$, depending on whether the excitation number is
even or odd. So the system has the corresponding even or odd parity.
It is easily proven that the wavefunction (6) with even and odd
parity is of the form
\begin{equation}
\left| \Psi _{\pm }\right\rangle =\ \left(
\begin{array}{l}
\sum_{n=0}^{N_{tr}}f_n\left| n\right\rangle _A \\
\pm \sum_{n=0}^{N_{tr}}f_n\left| n\right\rangle _B
\end{array}
\right)
\end{equation}
where $\Psi _{+}$ $\left( \Psi _{-}\right) $ is corresponding to
wavefunction with even(odd) parity. Inserting Eq. (13) into Eq. (3)
gives
\begin{equation}
\left[ \frac{\omega _r}2+\omega _r\left( m-\alpha ^2\right) \right]
f_m\mp \frac \Delta 2\sum_nD_{mn}f_n=Ef_m
\end{equation}
The level transition is only allowed between the even and odd parity, i.e. $%
E_i^{(\pm )}\Leftrightarrow E_j^{(\mp )}$. The transition between
the levels with the same parity is forbidden, $E_i^{(\pm
)}\nLeftrightarrow \ E_j^{(\pm )}$. The optical selection rules
related to the parity have been discussed
in the microwave-assisted transitions of superconducting quantum circuits%
\cite{liuyx,Deppe}.

\section{Results and discussions}

D\'{i}az et al diagonalize a restricted Hilbert space to a certain
number of photon states (in the Fock basis) and obtained fitted
parameters\cite {exp,pol}. The optimum fit of the experimental
results within the present theoretical scheme gives $I_p=515nA,
g/2\pi =0.82GHz, \omega _r/2\pi =8.13GHz, \Delta /h=4.25GHz$, very
close to their values.  The calculations in this paper are based on
these parameters, unless specified.

\begin{figure}[tbp]
\includegraphics[scale=0.6]{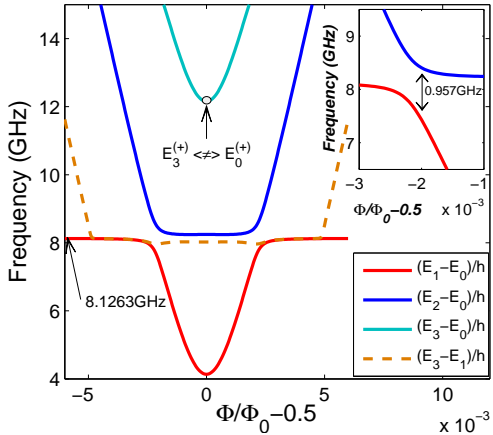}
\caption{ (Color online) Theoretical spectrum of the flux qubit
coupled to the LC resonator from numerical calculations.  }
\label{Spectrum}
\end{figure}

We plot the numerical results for the spectrum for $E_n\rightarrow
E_0$( $ n=1,2,$ and $3$)   in Fig.~\ref {Spectrum}. The experimental
three spectral lines are just corresponding to the transitions
between a few low energy levels, such as $ E_3\rightarrow E_0$
(upper), $E_2\rightarrow E_0$ (middle),
 and $E_1\rightarrow E_0$ (down). It is very interesting that our
theoretical results for the spectrum are in excellent agreement with
the experimental ones in Fig. 3 of Ref. \cite {exp}. Using these
fitted parameters, the energy splitting  on resonance $(E_2-E_1)/h$
obtained within the present  approach is around $ 0.957GHz$, just in
the scope of the experimental observation.

In Fig. 3 of Ref. \cite{exp}, a weakly visible spectrum line just
below the middle spectrum line was attributed to the thermally
excited qubit. We calculate the spectrum line for the transition
$E_3\rightarrow E_1$, as also list in Fig.~\ref{Spectrum} with a
yellow line. Interestingly, it is just in the location observed
experimentally shown in their Fig. 3. We believe that the state with
$E_1$ is just corresponding to the qubit excited thermally mentioned
in Ref. \cite{exp}.

We would like to mention here that the experimentally observed
spectrum lines have been explicitly related  to  the specified
energy level transitions in the JC model without RWA. Then the
comparison are easily performed.

\begin{figure}[tbp]
\includegraphics[scale=0.6]{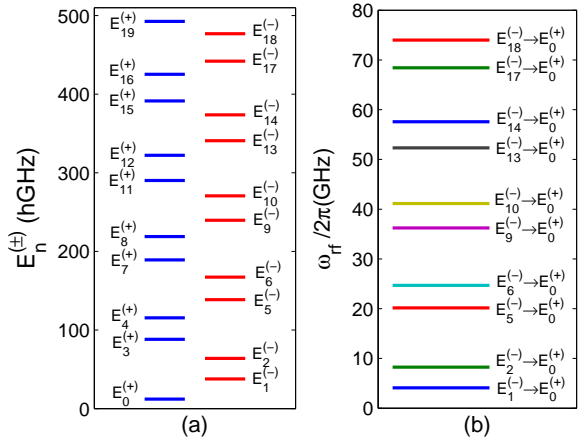}
\caption{ (Color online) (a) The energy levels $E_n^{(\pm )}$  and
(b) the first 10 spectrums $E_n^{(-)}\rightarrow E_0^{(+)}$ in the
symmetry point. } \label{energylevel}
\end{figure}

Next, we specially consider the case in  the symmetry point  .
Fig.~\ref{energylevel} (a) presents the energy levels from the
numerically exact calculations. It was suggested in Ref. \cite{exp}
that in the blue sideband spectral line \cite{Chiorescu} the minimum
vanishes since the qubit is in the symmetry point where it produces
no net flux and the transition is forbidden. In the symmetry point,
the transition from $E_3^{(+)}\rightarrow E_0^{(+)}$ is forbidden
due to the same parity, as shown in Fig.~\ref{energylevel}. This is
the reason that the upper spectrum line around the symmetry point of
Fig. 3 in Ref. \cite{exp} is almost invisible. It is perhaps just
the optical selection rules related to the parity makes the qubit to
produce no net flux. As also indicated in Fig.~\ref{energylevel},
the other two transitions between levels with the different parity
are allowed, so the intensities in the middle and down spectrum
lines in the symmetry point are nearly the same as in the whole
spectrum line.

Fig.~\ref{energylevel}(b) shows the first 10 spectrums $E_n^{(-)}\rightarrow
E_0^{(+)}$ theoretically. In the experimental accessible detection, one can
check the existence of these spectrums.

\begin{figure}[tbp]
\includegraphics[scale=0.7]{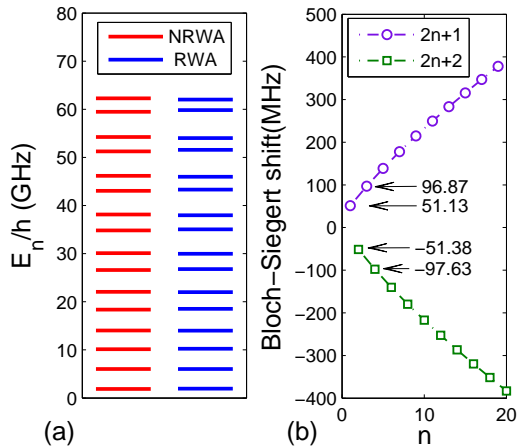}
\caption{ (Color online) (a) The energy levels $E_i^{(\pm )}$ for in
the symmetry point obtained within non-RWA (left panels) and RWA
(right panels). (b) Bloch-Siegert shift for the different level
transitions $E_i\rightarrow E_0$ in the symmetry point. }
\label{B_S_shift}
\end{figure}

We then turn to the Bloch-Siegert shift, which is just energy shift
of the level transition with the consideration of the
counter-rotating terms in the ultrastrong coupling regime. The
Bloch-Siegert shift of the level transitions $E_i\rightarrow E_0$ in
the symmetry point are exhibited in Fig. \ref{B_S_shift}. The
smallest Bloch-Siegert shift is around $50MHz$, nearly the same as
that measured in the experiment\cite{exp}. Note that some level
transitions $E_i\rightarrow E_0$ are forbidden in the symmetry point
due to the same parity, which are also presented here only for the
estimation of the magnitude of the Bloch-Siegert shift in the
corresponding spectrum. For the main spectrum $E_i\rightarrow E_0$,
the Bloch-Siegert shift becomes larger as $i$ increases, and its
sign changes alternatively with either $i$, as shown in Fig.
\ref{B_S_shift}(b).

\begin{figure}[tbp]
\includegraphics[scale=0.6]{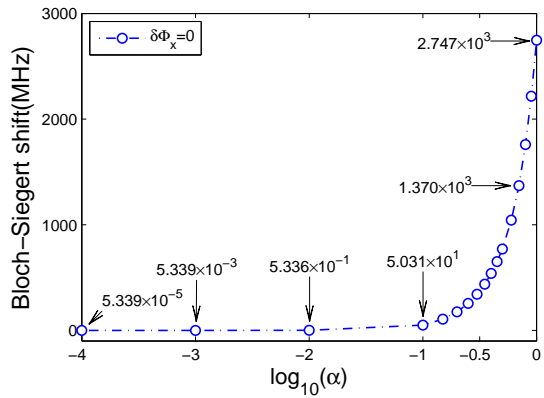}
\caption{ (Color online) Bloch-Siegert shift of the level
transitions $E_1\rightarrow E_0$ in the symmetry point versus
$\alpha=g/\omega_r$. } \label{shift_g}
\end{figure}

To show the effect of the qubit-cavity coupling strength on the
Bloch-Siegert shift, we fix all parameters fitted from experiments
except the coupling parameter $g$. The Bloch-Siegert shift of the
level transitions $E_1\rightarrow E_0$ in the symmetry point as a
function of the effecting coupling constant $\alpha=g/\omega_r$
defined in Eq. (5) are plotted in Fig. \ref{shift_g}. In the weak
coupling regime, say $\alpha\le0.01$, the Bloch-Siegert shift is so
small ( less than 1 $MHz$)  that it could not be distinguished from
the spectrum line. When $\alpha \ge 0.1$, the Bloch-Siegert shift
increases considerably with the coupling constant, and can reach the
regime of $GHz$. This observation is also of practical interest.
Recently, the coupling could easily be further enhanced in the
circuit QED \cite{Devoret,Bourassa,Niemczyk} where  $g$ is
comparable with $\omega_r$, i.e. $\alpha$ is in the order of
magnitude of $1.0$. If $\alpha>0.15$, the calculated Bloch-Siegert
shift is observed to exceed $80 MHz$, the  qubit line width at the
symmetry point around $4 GHz$, it is predicted that the
Bloch-Siegert shift could be clearly resolved experimentally in this
strong-coupling regime, like the Lamb shift \cite{Fragner}.

\section{conclusions}
In summary, by using extended bosonic coherent states, we solve the
Jaynes-Cummings model without RWA exactly in the numerical sense.
Within this technique, we can reproduce excellently the spectrum
measured in a recent experiments on an LC resonator magnetically
coupled to a superconducting qubit\cite{exp}, which was demonstrated
in the ultra-strong coupling regime. The Bloch-Siegert shift
$E_1\rightarrow E_0$ in the symmetry point is estimated to be
$50MHz$, very close to the experimental value. For the transition
between the higher excited state $i>1$ and the ground-state, the
magnitude of the Bloch-Siegert shift monotonously increases, but the
sign changes as $(-1)^{(i+1)}$. The considerable Bloch-Siegert shift
in turn demonstrate that the counter-rotating terms should be
considered. The effect of the qubit-cavity coupling strength on the
Bloch-Siegert shift is also investigated. It is predicted that the
Bloch-Siegert shift can  be distinguished experimentally for
$\alpha>0.15$. The present technique are more suited for the
stronger coupling regime, which experimental realizations may appear
in the near future\cite{Devoret,Bourassa,Niemczyk}.

\section{Acknowledgements}
The authors acknowledges useful discussions with P. Forn-D\'{i}az.
This work was supported by National Natural Science Foundation of
China under Grant Nos. 10974180, National Basic Research Program of
China (Grant Nos. 2011CB605903 and 2009CB929104).

\end{document}